\documentclass[preprint
    ,final            
  ]
  {aipproc}

\layoutstyle{6x9}

\begin{document}

\begin{flushright}
Caltech MAP-325\\
CALT-68-2608
\end{flushright}

\title{Constraints on Muon Decay Parameters from Neutrino Mass}

\classification{12.15.-y, 13.35.Bv, 14.60.St}
\keywords      {muon decay, neutrino mass, electroweak interactions}

\author{Rebecca J. Erwin}{
  address={California Institute of Technology\\ Pasadena, CA 91125}
}

\author{Jennifer Kile}{
   address={California Institute of Technology\\ Pasadena, CA 91125}
}

\author{Michael J. Ramsey-Musolf}{
  address={California Institute of Technology\\ Pasadena, CA 91125}
}

\author{Peng Wang}{
  address={California Institute of Technology\\ Pasadena, CA 91125}
}

\begin{abstract}

We derive model-independent constraints on chirality-changing terms in the muon decay Lagrangian using limits on neutrino mass.  We consider all dimension-six operators invariant under the gauge symmetry of the Standard Model which contribute to either a Dirac neutrino mass or muon decay.  Taking an upper limit on neutrino mass of $1$eV, we derive limits on the contributions of chirality-changing operators to the Michel parameters four orders of magnitude tighter than the current experimental constraints.  We also identify two operators which, due to their flavor structure, are not constrained by neutrino mass.  If near-future experiments find contributions to muon decay from these operators, it could indicate interesting flavor structure in physics beyond the SM.

\end{abstract}

\maketitle


\section{Introduction}

In this paper (for a more detailed discussion see \cite{Erwin}), we use the current limits on neutrino mass of $\sim 1$eV \cite{Lobashev:1999tp,Weinheimer:1999tn} to constrain the contributions of physics beyond the Standard Model to chirality-changing terms in the muon decay Lagrangian
\begin{equation}
{\cal L}^{\mu-\rm decay} = \frac{4 G_\mu}{\sqrt{2}}\ \sum_{\gamma,\, \epsilon,\, \mu} \ g^\gamma_{\epsilon\mu}\, 
\ {\bar e}_\epsilon \Gamma^\gamma \nu {\bar\nu} \Gamma_\gamma \mu_\mu~. \end{equation}
We consider the contributions of all $SU(2)\times U(1)$-invariant dimension-6 operators 
\begin{equation}
{\cal L}_{\rm eff}^{(6)} = \sum_{j} \frac{C_j^6(\mu)}{\Lambda^{2}}\, O_j^{(6)}(\mu) \  + {\rm h.c.}
\label{eq:lagr}
\end{equation}
to muon decay.  Some terms in ${\cal L}_{\rm eff}^{(6)}$ mix at 1-loop order with the 4D and 6D neutrino Dirac mass operators 
\begin{equation}
O^{(4)}_{M,\, AD} \equiv {\bar L}^A {\tilde\phi} \nu_R^D  \quad \mbox{and} \quad O^{(6)}_{M,\, AD} \equiv \bar{L}^A\widetilde{\phi}\nu_{R}^D(\phi^{+}\phi) ~ \nonumber
\end{equation}
where  $A,B,C,D$ are flavor indices and $\widetilde{\phi}=i\tau^2\phi$.  After electroweak symmetry breaking, these neutrino mass operators give contributions to $m_{\nu}$ of 
\begin{equation}
\delta m_\nu^{(4)AD} = \frac{-v}{\sqrt{2}}C^4_{M,\, AD}\, (v) \quad \mbox{and} \quad 
\delta m_\nu^{(6)AD}  =  \frac{-v^3}{2\sqrt{2}\Lambda^2} C^6_{M,\, AD}(v)~. 
\end{equation}
Thus, $m_\nu$ can give constraints on the terms in Eq. (\ref{eq:lagr}) which mix into $O^{(4)}_{M,AD}$ and $O^{(6)}_{M,AD}$.

\section{Calculation}

We list in Table \ref{tab:ops} the linearly independent operators contributing to $m_{\nu}$, with their contributions to $\mu$ decay.  Only $O^{(6)}_{{\tilde V},\, AD}$ and $O^{(6)}_{F,\, ABCD}$ are important for $\mu$ decay.  $O_{B,\, AD}^{(6)}$ and $O_{W,\, AD}^{(6)}$ are suppressed due to the derivative on the guage boson.  All other 6D operators which contribute significantly to $\mu$ decay affect only $g_{\epsilon \mu}$ with $\epsilon=\mu$.  
\begin{table}
\label{tab:ops}
\caption{6D operators which contribute to $m_{\nu}$ and their contributions to the $g^{\gamma}_{\epsilon \mu}$.} 
\begin{tabular}{lc}
 \tablehead{1}{l}{b}{Operator}
& \tablehead{1}{c}{b}{$\mu$ Decay Contribution}\\ 
\hline
$O_{B,\, AD}^{(6)}  =  g_{1}(\bar{L}^A\sigma^{\mu\nu}\widetilde{\phi})\nu_{R}^D B_{\mu\nu}$ & $(m_{\mu}/v)^2$ suppressed \\ 
$O_{W,\, AD}^{(6)} =  g_{2}(\bar{L}^A \sigma^{\mu\nu}\tau^{a}\widetilde{\phi})\nu_{R}^DW_{\mu\nu}^{a}$ & $(m_{\mu}/v)^2$ suppressed \\ 
$O^{(6)}_{M,\, AD} = (\bar{L}^A\widetilde{\phi}\nu_{R}^D)(\phi^{+}\phi)$ & None \\ 
$O^{(6)}_{{\tilde V},\, AD}  = i(\bar{\ell}_{R}^A\gamma^{\mu}\nu_{R}^D)(\phi^{+}D_{\mu}\widetilde{\phi})$ & $g^V_{RL,LR}$ \\ 
$O^{(6)}_{F,\, ABCD} = \epsilon^{ij}\bar{L}_{i}^A\ell_{R}^C\bar{L}_{j}^B\nu_{R}^D $ & $g^{S,T}_{RL,LR}$ \\ 
\hline
\end{tabular}
\end{table}

We calculate the mixing of the $O^{(6)}_j$ into $O^{(4)}_{M,\, AD}$ and $O^{(6)}_{M,\, AD}$.  We obtain order-of-magnitude estimates for their contributions to $C^4_{M,AD}$: 
\begin{eqnarray}
\nonumber
O_{B,\, AD}^{(6)} & \rightarrow & C^4_{M,\, AD} \sim \frac{\alpha}{4\pi \cos^2\theta_W} C^6_{B,\, AD} \\
\nonumber
O_{W,\, AD}^{(6)} & \rightarrow & C^4_{M,\, AD} \sim \frac{3\alpha}{4\pi \sin^2\theta_W} C^6_{W,\, AD}\\
  \nonumber 
O^{(6)}_{{\tilde V},\, AD} & \rightarrow & C^4_{M,\, AD}\sim \frac{f_{AA}}{16\pi^2} C^6_{{\tilde V},\, AD}\\
\nonumber
O^{(6)}_{F,\, BABD} & \rightarrow & C^4_{M,\, AD}\sim \frac{f_{BB}}{4\pi^2} C^6_{F,\, BABD}\\
\nonumber
O^{(6)}_{F,\, ABBD} & \rightarrow & C^4_{M,\, AD}\sim \frac{f_{BB}}{16\pi^2} C^6_{F,\, ABBD}
\end{eqnarray}

To obtain the contributions of $O^{(6)}_{j}$ to $C^6_{M,AD}$, we calculate the mixing amongst the 6D operators and solve the renormalization group equations for the coefficients $C^6_j(v)$.  The resulting limits are weaker by $\sim (v/\Lambda)^2$ than those from mixing into $O^{(4)}_{M,AD}$. \\





\section{Results and Conclusions}

Table \ref{tab:gconstraints} gives our upper limits on the $g^{S,V,T}_{LR,RL}$.  These bounds are $\sim4$ orders of magnitude stronger than a recent global fit to the experimental data \cite{Gagliardi:2005fg} and $\sim2$ orders of magnitude stronger than the results of a two-loop analysis \cite{Prezeau:2004md} of constraints from neutrino mass.  We note that these limits are model-independent, but could be evaded by fine-tuning.

\begin{table}
\caption{Upper bounds on the contributions of the $O^{(6)}_{j}$  to the $|g^{S,V,T}_{LR,RL}|$.}
\label{tab:gconstraints}
\begin{tabular}{@{}lcccccc @{}}
\\
 \tablehead{1}{l}{b}{$O^{(6)}_j$}
  & \tablehead{1}{c}{b}{$|g^S_{LR}|$}
  & \tablehead{1}{c}{b}{$|g^T_{LR}|$}
  & \tablehead{1}{c}{b}{$|g^S_{RL}|$}
  & \tablehead{1}{c}{b}{$|g^T_{RL}|$}
  & \tablehead{1}{c}{b}{$|g^V_{LR}|$}
  & \tablehead{1}{c}{b}{$|g^V_{RL}|$}\\
\hline
$O^{(6)}_{F,\, e\mu\mu D}$ & $4\times 10^{-7}$ & $2\times 10^{-7}$ & - & - & - & - \\ 
$O^{(6)}_{F,\, \mu e\mu D}$ & $4\times 10^{-7}$ & - & - & - & - & - \\ 
$O^{(6)}_{F,\, \mu eeD}$  & - & -&  $8\times 10^{-5}$ & $4\times 10^{-5}$ & - & - \\ 
$O^{(6)}_{F,\, e\mu eD}$ & - & - & $8\times 10^{-5}$ & - & - & - \\ 
$O^{(6)}_{{\tilde V},\, \mu D}$ & - & - & - & - & $8\times 10^{-7}$ & - \\ 
$O^{(6)}_{{\tilde V},\, eD}$ & - & - & - & - & - & $2 \times 10^{-4}$  \\  \hline
\end{tabular}
\end{table}

Two operators not shown in the table, $O^{(6)}_{F,\, ee\mu D}$ and $O^{(6)}_{F,\, \mu\mu eD}$, contribute to $g^{S,T}_{LR}$ and $g^{S,T}_{RL}$, respectively, but are not constrained by $m_{\nu}$.  However, as they differ from other $O^{(6)}_{F,\, ABCD}$ only by flavor, we naively expect their contributions to $g^{S,T}_{RL,LR}$ to be similar to those of the other $O^{(6)}_{F,\, ABCD}$.  An observed large contribution to $g^{S,T}_{RL,LR}$ could be an indication of new physics with interesting flavor structure.  These operators could be relevant if current measurements of the Michel parameter \cite{Michel,Kinoshita} $\rho$ by TWIST \cite{tribble} give a value disagreeing with the Standard Model.



\begin{theacknowledgments}
The authors thank N. Bell, V. Cirigliano, M. Gorshteyn, P. Vogel, and M. Wise for helpful discussions.  J. Kile thanks the organizers of CIPANP 2006 for the opportunity to present these results.  This work was supported in part under U.S. Department of Energy contracts FG02-05ER41361 and DE-FG03-92ER40701 and National Science Foundation award PHY-0071856.
\end{theacknowledgments}





\begin{thebibliography}{99}

\bibitem{Erwin}
  R.~J.~Erwin, J.~Kile, M.~J.~Ramsey-Musolf and P.~Wang,
  [arXiv:hep-ph/0602240].
\bibitem{Lobashev:1999tp}
  V.~M.~Lobashev {\it et al.},
  Phys.\ Lett.\ B {\bf 460}, 227 (1999).
  
\bibitem{Weinheimer:1999tn}
  C.~Weinheimer {\it et al.},
  Phys.\ Lett.\ B {\bf 460}, 219 (1999).
  
\bibitem{Gagliardi:2005fg}
  C.~A.~Gagliardi, R.~E.~Tribble and N.~J.~Williams,
  Phys.\ Rev.\ D {\bf 72}, 073002 (2005)
  [arXiv:hep-ph/0509069].

\bibitem{Prezeau:2004md}
  G.~Prezeau and A.~Kurylov,
  Phys.\ Rev.\ Lett.\  {\bf 95}, 101802 (2005)
  [arXiv:hep-ph/0409193].





  

\bibitem{Michel} L. Michel, 
Proc.\ Roy.\ Soc.\ Lond.\  {\bf A63}, 514 (1950); C. Bouchiat and L. Michel,
Phys.\ Rev.\  {\bf 106}, 170 (1957).

\bibitem{Kinoshita} T. Kinoshita and A. Sirlin, 
Phys.\ Rev.\  {\bf 108}, 844 (1957).



\bibitem{tribble} R. Tribble and G. Marshall, private communication. See also Hu, J to appear in the proceedings of the PANIC05 conference (Sante Fe, N.M.), 2005.











  %



  %

  %




.


\end{thebibliography}


\end{document}